\documentclass[apj]{aastex}
\usepackage{pslatex}
\usepackage{graphicx}
\shorttitle{X-ray surface brightness profiles of AGN}
\shortauthors{Chatterjee et al.}
\begin{document}
\title{X-ray Surface Brightness Profiles of Active Galactic Nuclei in the Extended Groth Strip: Implications for AGN Feedback} 

\author{Suchetana Chatterjee\altaffilmark{1,2,3}, Jeffrey A.\ Newman\altaffilmark{1,4}, Tesla Jeltema\altaffilmark{5}, Adam D.\ Myers\altaffilmark{2}, James Aird\altaffilmark{6}, Alison L.\ Coil\altaffilmark{7}, Michael Cooper\altaffilmark{8}, Alexis Finoguenov\altaffilmark{9,10}, Elise Laird\altaffilmark{11}, Antonio Montero-Dorta\altaffilmark{12}, Kirpal Nandra\altaffilmark{13}, Christopher Willmer\altaffilmark{14}, Renbin Yan\altaffilmark{15} }
\altaffiltext{1}{Department of Physics and Astronomy, University of Pittsburgh, Pittsburgh, PA 15260, USA}
\altaffiltext{2}{Department of Physics and Astronomy, University of Wyoming, Laramie, WY 82072, USA}
\altaffiltext{3}{Yale Center for Astronomy and Astrophysics, Department of Physics, Yale University, New Haven, CT 06520, USA}
\altaffiltext{4}{PITT-PACC, University of Pittsburgh, Pittsburgh, PA 15260, USA}
\altaffiltext{5}{Department of Physics, University of California, Santa Cruz, CA 95064, USA}
\altaffiltext{6}{Department of Physics, Durham University, Durham DH13LE, UK}
\altaffiltext{7}{Department of Physics, University of California, San Diego, CA 92093, USA}
\altaffiltext{8}{Department of Physics and Astronomy, University of California, Irvine, CA 92697, USA}
\altaffiltext{9}{Department of Physics, University of Helsinki, Helsinki, Finland}
\altaffiltext{10}{Center for Space Science Technology, University of Maryland Baltimore County, Baltimore, MD 21250, USA}
\altaffiltext{11}{Astrophysics Group, Imperial College London, Blackett Laboratory, Prince Consort Road, London SW7 2AZ, UK}
\altaffiltext{12}{Department of Physics and Astronomy, University of Utah, Salt Lake City, UT 84112, USA}
\altaffiltext{13}{Max Planck  Institut f\"{u}r Extraterrestrische  Physik, Giessenbachstra\ss e, 85748 Garching, Germany}
\altaffiltext{14}{Steward Observatory, University of Arizona, Tucson, AZ, 85721, USA}
\altaffiltext{15}{Department of Physics and Astronomy, University of Kentucky, Lexington, KY 40506, USA}

\begin{abstract}
Using data from the All Wavelength Extended Groth Strip International Survey (AEGIS) we statistically detect the extended X-ray emission in the interstellar medium (ISM)/intra-cluster medium (ICM) in both active and normal galaxies at $0.3 \leq z \leq 1.3$. For both active galactic nuclei (AGN) host galaxy and normal galaxy samples that are matched in restframe color, luminosity, and redshift distribution, we tentatively detect excess X-ray emission at scales of 1--$10\arcsec$ at a few $\sigma$ significance in the surface brightness profiles. The exact significance of this detection is sensitive to the true characterization of {\it Chandra's} point spread function. The observed excess in the surface brightness profiles is suggestive of lower extended emission in AGN hosts compared to normal galaxies. This is qualitatively similar to theoretical predictions of the X-ray surface brightness profile from AGN feedback models, where feedback from AGN is likely to evacuate the gas from the center of the galaxy/cluster. We propose that AGN that are intrinsically under-luminous in X-rays, but have equivalent bolometric luminosities to our sources will be the ideal sample to study more robustly the effect of AGN feedback on diffuse ISM/ICM gas.    
\end{abstract}

\keywords{galaxies: active, galaxies:ISM, ICM, AGN:general, Xrays:ISM, galaxies}

\section{Introduction}
Several lines of evidence suggest that energy input from active galactic nuclei (commonly known as AGN feedback) can have substantial effects on the formation and evolution of galaxies. For example, the observed correlation between black hole mass-bulge mass \citep[e.g.,][]{gebhardtetal00, m&f01, tremaineetal02} strongly implies a connection between galaxy formation and black hole growth. The observed lack of expected cooling flows in galaxy clusters and the exponential cut-off at the bright end of the galaxy luminosity function have also been linked with AGN feedback \citep[e.g.,][]{p&f06, crotonetal06}.

Effects of AGN feedback have been directly observed in groups and clusters using multi-wavelength data. For example, AGN residing in cluster centers have supermassive black holes, which accrete matter from the intra-cluster medium (ICM), releasing tremendous amounts of energy in radiation and/or outflows. This has been observed with X-rays in cluster cores \citep[e.g.,][]{m&n07}. With the emergence of {\it Chandra} and XMM-Newton, the evidence that the central radio sources in groups and clusters have a profound, persistent effect on the ICM has been strongly established. It has been shown that the deficits in the X-ray emission in clusters (X-ray cavities) are spatially coincident with regions of high synchrotron emission \citep[e.g.,][]{birzanetal04, nulsenetal05, d&f06, m&n07, gittietal12}. Galaxy groups have shallower potential wells and hence smaller intrinsic thermal energy. Thus AGN outbursts have a large impact on the intragroup medium \citep[e.g.,][]{giodinietal10}. The connection between X-ray cavities and AGN activity in the intragroup medium has been frequently studied \citep[e.g.,][]{johnsonetal09, osullivanetal10, dongetal10, randalletal11}.

\begin{figure}[t]
\epsscale{1.00}
\plotone{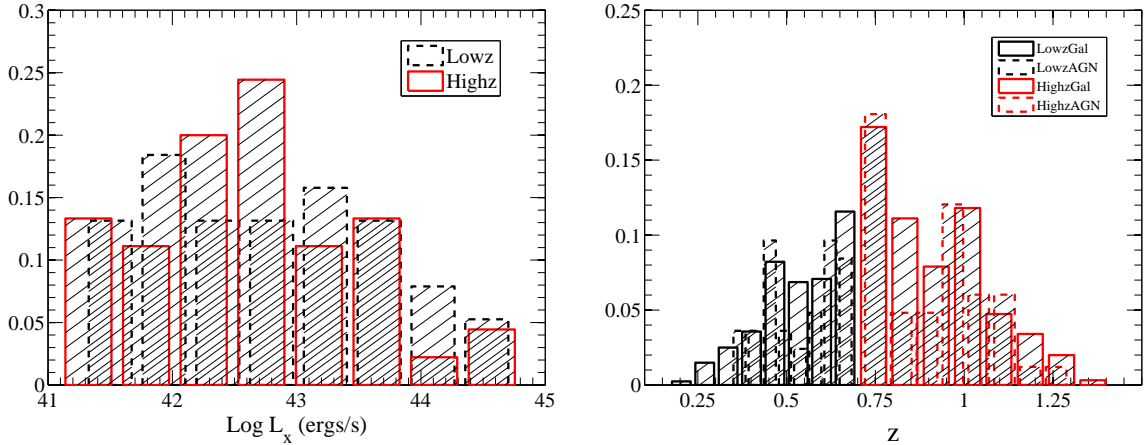}
\caption{Distribution of soft X-ray luminosity (left-hand panel) and redshift (right-hand panel) for our X-ray AGN sources (L09 sample). We present the luminosity and redshift distributions of our {\it Lowz} ($0.3 \leq z \le 0.7$) and {\it Highz} ($0.7 \leq z \le 1.3$) samples. In the left-hand panel the red solid and the black dashed histograms show the distribution of X-ray luminosities for the {\it Highz} and the {\it Lowz} sample, respectively. The solid and the dashed histograms in the right-hand panel depict the redshift distributions of the galaxy control sample and the AGN sample, respectively. Red and black refer to the {\it Highz} and the {\it Lowz} sample, respectively.  The distributions in both the panels are normalized by the total number of objects in each sample. See \S 2 for more details. The bin sizes and labels are slightly different for each sample.} \label{fig-1}
\end{figure}

Several theoretical models relating AGN activity to galaxy evolution have been proposed in this context. In many of these models AGN feedback is introduced in the form of thermal energy feedback which naturally explains the $M-\sigma$ relation and exponential cut-off of the bright end of the galaxy luminosity function \citep[e.g.,][] {k&h00, c&o01, w&l03, marconietal04, shankaretal04, dimatteoetal05, cattaneoetal06, crotonetal06, hopkinsetal06, lapietal06, b&s09,  teyssieretal11, johanssonetal08}. Some models do include momentum feedback \citep[e.g.,][]{sijackietal07, c&o07, novaketal11, gasparietal11,  gasparietal12, choietal12, choietal13} but the scales and physical processes vary widely between them. 

The effect of feedback on several observables has been explored in the literature, including the $L_{x}-T$ relation in galaxy clusters and groups \citep[e.g.][]{a&e99, n&r02, s&o04, p&f06, thackeretal09, puchweinetal10}, Sunyaev-Zeldovich (SZ; \citealt{s&z72}) profiles \citep[e.g.,][] {bhattacharyaetal08, chatterjeeetal08, chatterjeeetal10}, SZ power spectrum \citep[e.g.,][]{c&k07, scannapiecoetal08, battagliaetal10}, and star-formation properties of galaxies \citep[e.g.,][] {dimatteoetal05, cattaneoetal06, crotonetal06, hopkinsetal06, schawinskietal07}. Recently, theoretical studies have attempted to quantify the effect of AGN feedback on the properties of the X-ray gas in the ISM of elliptical galaxies \citep[e.g.,][]{pelligrinietal12, gasparietal12, choietal13}. Motivated in part by these studies we now investigate the impact of AGN on their large-scale environments by examining the properties of the diffuse X-ray emitting gas in AGN host galaxies at high redshift ($z \sim 0.8$). 

By using a large sample of active and normal galaxies which have identical optical properties we statistically compare their diffuse X-ray emission to evaluate the impact of AGN activity on the ISM/ICM gas. We employ X-ray data from the All Wavelength Extended Groth Strip International Survey (AEGIS) project \citep{davisetal07} to obtain the X-ray surface brightness profiles of our samples. The AEGIS-X survey \citep{nandraetal07} optimizes the balance between depth and sky area covering a sky region of $0.67\,{\rm deg^{2}}$ in the energy range 0.5--7\,kev. The survey region has been scanned at many wavelengths from radio to X-rays. The wide field of view, and the broad coverage in redshift, along with multi-wavelength observations, makes AEGIS a premier dataset for studying AGN co-evolution. We perform a stacking analysis of X-ray maps of AGN and normal galaxies from the AEGIS-X survey and compare their mean X-ray surface brightness profiles to investigate the effect of AGN on the ISM/ICM. 

Our paper is organized as follows. In \S 2 we give a brief description of our datasets. In \S 3 we describe the methodology. We present our results in \S 4. We finally discuss our results and summarize our conclusions in \S 5 and \S 6. Throughout the paper we assume a spatially flat, $\Lambda$CDM cosmology: $\Omega_{m}=0.28$, $\Omega_{\Lambda}=0.72$, $\Omega_{b}=0.04$, and $h=0.71$.

\begin{figure*}[t]
\begin{center}
\begin{tabular}{c}
        \resizebox{40mm}{!}{\includegraphics{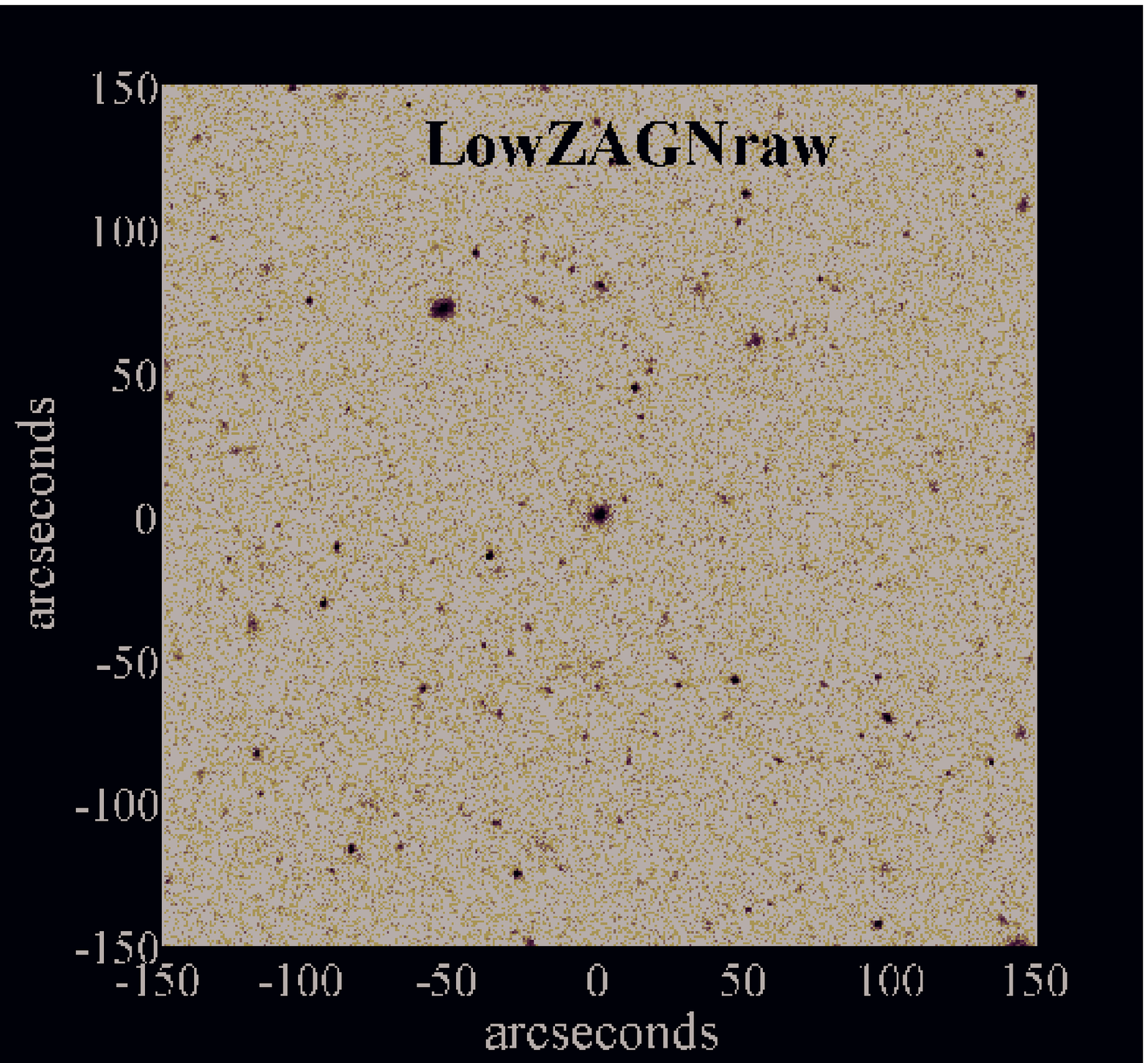}}
       \resizebox{40mm}{!}{\includegraphics{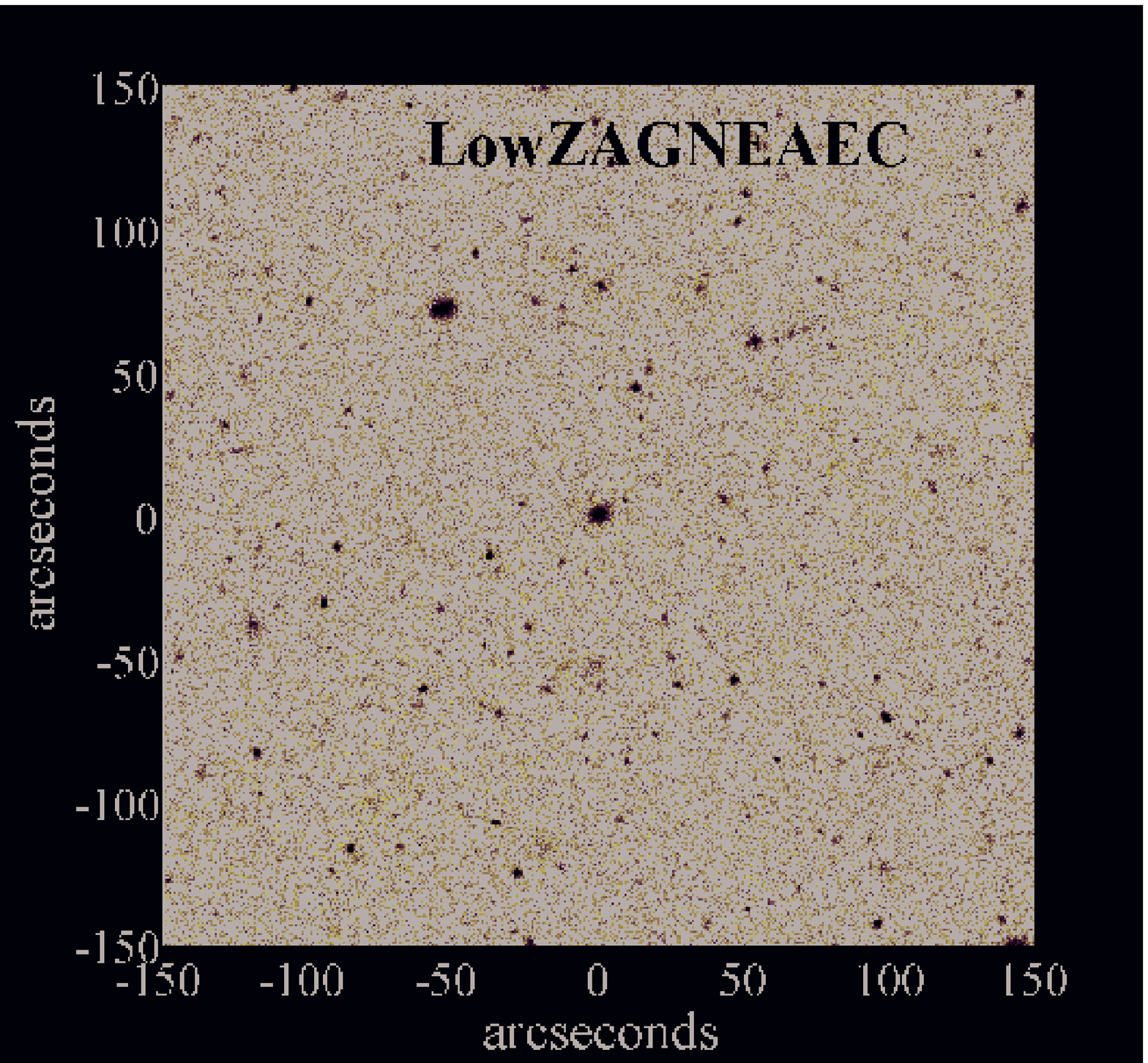}}
      \resizebox{40mm}{!}{\includegraphics{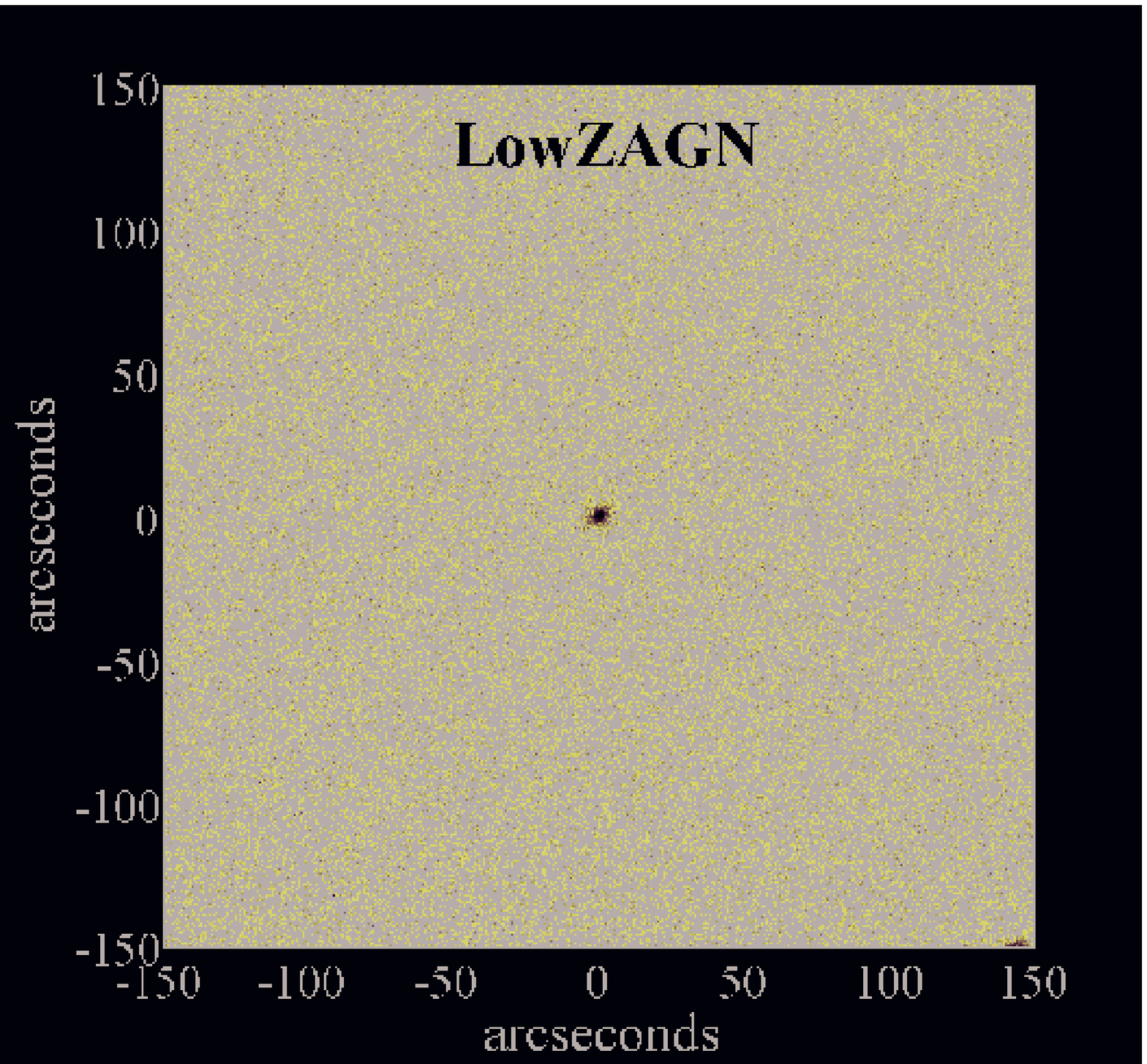}}
       \resizebox{40mm}{!}{\includegraphics{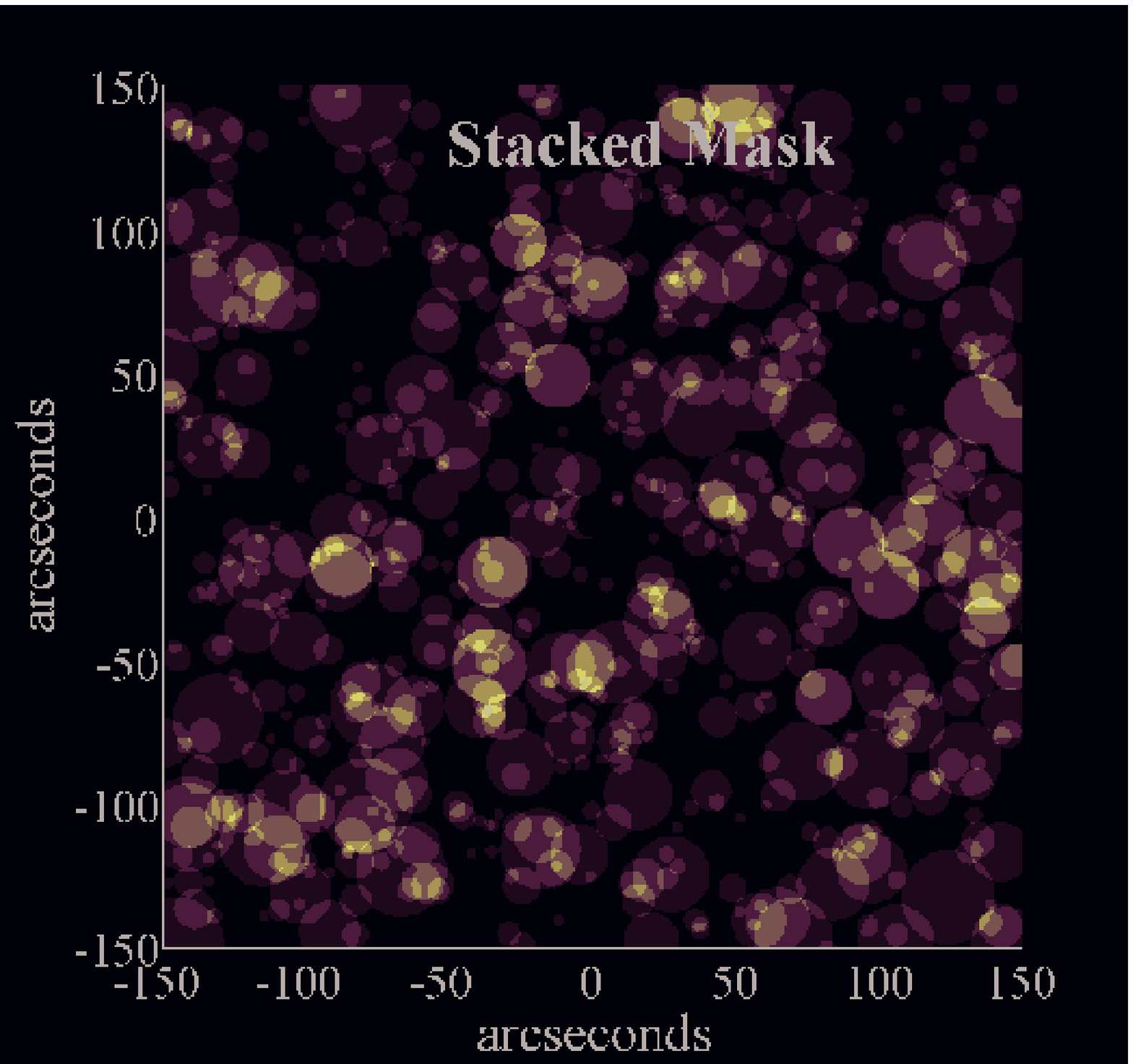}}\\
      
\end{tabular}
 \caption{Methodology for constructing our stacked maps. All the maps are for the soft band corresponding to the L09 {\it Lowz} sample. The method is described in \S 3. The left-most panel shows the stacked map of the raw event files. This map is not corrected for exposure time or effective area, and is not masked for point sources. The middle-left panel shows the same map after correcting for exposure time and effective area. The point sources are still not masked in this map. We call this the EAEC map. We emphasize that these two maps are for illustrative purposes and we do not use them in our actual analyses---instead, we correct the 8 EGS event files with the corresponding effective area-exposure time maps. See \S 3 for more discussion. The middle-right panel shows the exposure time effective area corrected map, where the point sources are masked using the method described in L09. The right-most panel shows the stacked mask map. Our final maps are constructed by dividing the middle-right panel by the right-most panel. }\label{fig-2}
\end{center}
\end{figure*}

\section{Data Sets}

For the purpose of our analysis we use data from the AEGIS-X survey (\citealt{lairdetal09}; L09 hereafter) and the DEEP2 Galaxy Redshift Survey \citep{davisetal03, newmanetal13}. In this section we will provide a brief description of our datasets.

\subsection{X-ray AGN sample}

The AEGIS-X survey consists of 8 deep {\it Chandra} ACIS-I pointings, each with a total integration time of about 200\,ks covering an area of $0.67 {\rm\,deg^{2}}$. Details of data reduction are discussed in L09. The individual observations are merged into a single event file and images are constructed in four energy bands 0.5--7.0\,keV (full), 0.5--2.0\,keV (soft), 2.0--7.0\,keV (hard) and 4.0--7.0\,keV (ultra-hard). The limiting flux in each of these bands is estimated to be $2.37 \times 10^{-16}$, $5.31  \times 10^{-17}$,  $3.76 \times  10^{-16}$ and $\rm  6.24 \times 10^{-16}  \,  erg \,  s^{-1}  \,  cm^{-2}$, respectively (L09). We use the soft band for our analysis. 

A point source catalog of the Extended Groth Strip (EGS) field has been provided in L09. The catalog consists of a total of $1325$ band-merged sources with a Poisson probability limit of $4\times 10^{-6}$. The source detection algorithm is described in \citet{nandraetal05}. The basic technique is based on pre-detection with a low signal-to-noise threshold and follow-up aperture extraction of the photon counts to determine the detection significance of the source. The X-ray catalogs were matched to the DEEP2 optical photometry catalog \citep{coiletal04} to account for positional offsets.   

To select our AGN sample from the point source catalog of L09 we applied the following cuts to the dataset. We first applied a cut on the soft X-ray luminosities of our sources and selected objects with soft X-ray luminosities $\geq 10^{41}$ ergs s$^{-1}$. The luminosity cut should retain only sources with X-ray emission due to AGN. We then applied a redshift cut to our sample ($0.3 \leq z \leq 1.3$) to be consistent with the bulk of the range covered by galaxies with secure redshifts from the DEEP2 Galaxy Redshift Survey. 

Finally we restricted our analysis to the sources that have good quality spectroscopic redshift measurements from DEEP 2. For our AGN sources we have additional spectroscopic redshifts, as described in \citet{coiletal09}. This is required to construct a reliable control sample of galaxies for reasons described in \S 2.2. Out of the $1325$ sources $477$ of them have DEEP2 counterparts. $194$ of them have redshift measurements. The high quality spectroscopic redshift criterion, and the subsequent redshift and luminosity cuts reduce our sample size to $96$ AGN. For our surface brightness analysis we construct two redshift subsamples which we will refer to as {\it Lowz} ($0.3 \leq z \leq 0.7$: 51 sources) and {\it Highz} ($0.7 \le z \leq 1.3$: 45 sources). The cut that split the two samples by redshift was selected at the median of the distribution. In Fig.\ 1 we show the luminosity (left panel) and redshift (right panel) distributions of our X-ray AGN sources. In both panels, the distributions are normalized by the total number of objects in each subsample.

\subsection{Galaxy Control Sample}

It is well known that ISM properties should depend on both the stellar mass of a galaxy and its color (e.g., cold gas fractions and virial temperatures are related to these properties of a galaxy). Hence, if we want to identify impacts of AGN activity on galaxies, we must compare samples that are matched in stellar mass and color. Luminosity and color provide excellent proxies for stellar mass and color \citep[e.g.,][]{b&dejong01}. We also expect X-ray properties to depend on redshift---since galaxy ages, luminosity distance, angular diameter distance, and surface brightness dimming will all evolve with redshift. We therefore construct a set of galaxies which matches the distribution of our AGN in color, luminosity, and redshift. We also note that the large scale environments of galaxies and AGN will affect the ISM emission significantly. Studies show that there is no statistically significant difference between the environments of AGN and of other galaxies that are matched in color and luminosity \citep[e.g.,][]{georgakakisetal07, monterodortaetal09}. Thus, matching our samples of galaxies and AGN
should minimize environmental differences. Hence, if AGN feedback has no short-term effects, the ISM emission from both our AGN and our control galaxy sample should be identical.

The galaxy control samples for the L09 objects were constructed using the method described in \citet{cooperetal09}. The objects are matched to the AGN sample within the 3 dimensional parameter space of $B$-band absolute magnitude, restframe $U-B$ color, and redshift. The initial control sample consisted of 5000 galaxies, of which 2982 galaxies are unique. Some of the galaxies are stacked multiple times to ensure matched distributions of all parameters (i.e., color, redshift and luminosity). For extracting the surface brightness profiles we divide the galaxy sample into two redshift subsamples, using the same cuts as for the AGN redshift subsamples. The redshift distribution of the comparison sample is shown as the solid lines in the right-hand panel of Fig.\ 1.
 
\begin{figure}[t]
\epsscale{1.00}
\plotone{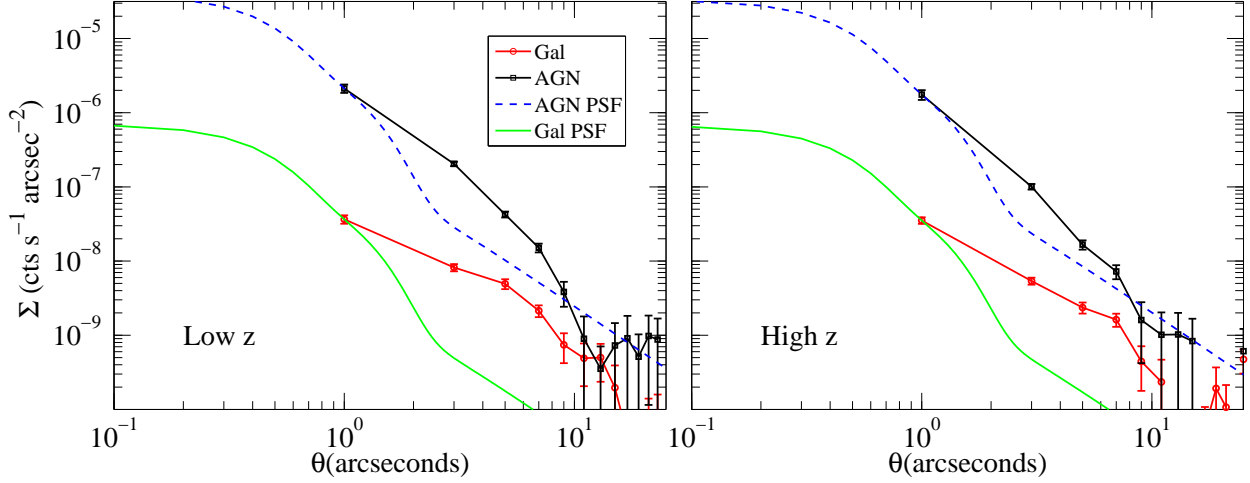}
\caption{Surface brightness profiles of AGN and galaxies in {\rm cts\,s}$^{-1}$arcsecond$^{-2}$. In both panels black squares correspond to the AGN sample and red circles correspond to the galaxy sample. The blue dashed line represents the average PSF model for our AGN sources and the green solid line is the approximate PSF profile of galaxies. The method for constructing the average PSF is discussed in \S 3. We normalize our PSF profiles to the peak of the AGN and galaxy emission. We detect excess emission in both galaxies and AGN over the range
 2--10 \arcsec. The left-hand and right-hand panels represent the {\it Lowz} and {\it Highz} subsamples, respectively. See Fig.\ 4  and \S 4 for a discussion of the mean difference profiles.} \label{fig-3}
\end{figure}

\begin{figure}[t]
\epsscale{0.80}
\plotone{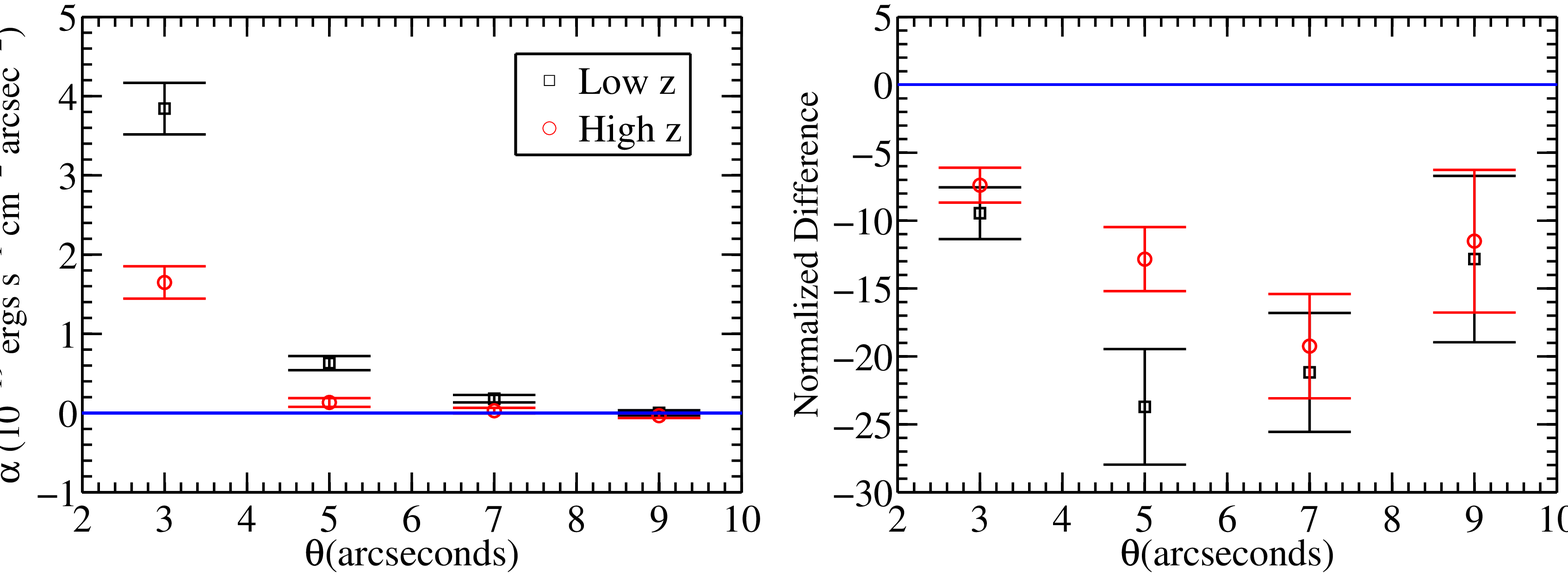}
\caption{The difference in the photon flux (quantified as $\alpha$ in Eq.\ 1) between the AGN and galaxy profiles as a function of angular scale. Positive (negative) numbers imply an excess (deficit) of X-ray flux in the AGN case. Black squares and red open circles represent $\alpha$ for the {\it Lowz} and {\it Highz} samples, respectively. The solid blue line represents $\alpha =0$. As noted in \S 4, a small number of our AGN and galaxy samples have a 95\% EER that covers the range 5--8\arcsec, which can contaminate the emission in Fig.\ 3 and contribute to a small amount of the excess seen in the left-hand panel. In the right-hand panel we plot $\alpha$, but now normalized to the corresponding PSFs of AGN and galaxies. We see more excess emission in galaxies compared to AGN when the difference is normalized by the PSFs. This is suggestive of less extended emission in AGN host galaxies as compared to galaxies that do not host an AGN. Our result is qualitatively similar to the ISM/ICM surface brightness profiles predicted from AGN feedback simulations where AGN feedback evacuates gas from the center of galaxies.} \label{fig-4}
\end{figure}

\section{Methodology}

Our measurements depend on the source stacking technique. In Fig.\ 2 we summarize the main components of our data stacking methodology using the example of {\it Lowz} L09 sources. The event files, the exposure time maps and the effective area maps are provided in L09. We first correct our event files using the exposure time-effective area maps. We call these effective area-exposure corrected (EAEC hereafter) maps. The gaps and singularities in the EAEC maps are corrected by assigning zero counts to those pixels where the EAEC maps have singularities (due to zero exposure and/or effective area). We then identify sources in the EAEC maps and select a $5\times 5$ square arcminute region around each source and sum them to construct the stacked image. 

Since we are interested in extended emission we mask the point sources in our stacked maps. We identify the point sources in each map from the point source catalog presented in L09. We calculate the point spread function (PSF) for each point source using the technique described in L09. The {\it getpsf} routine, provided by L09, is used to obtain the PSFs. We then mask the point sources using the ellipse corresponding to the $95\%$ encircled energy radius (EER). We mask the region which encompasses a circular area with a radius $1.5$ times that of the semi-major axis of the $95\%$ encircled energy ellipse of a particular point source. This allows us to create a conservative mask for each point source. We repeated our analyses with mask sizes of $1.0$ and $3.0$ times the $95 \%$ EER. We find that our results do not depend on the choice of our mask size and results shown in Figs.\ 3 and 4 are statistically identical in each case, although for more conservative masks the flux differences (Fig.\ 4) become increasingly insignificant. The results are similar to our fiducial mask sizes when we use a masking radius of $1.0$ times the $95\%$EER.

We also construct a stacked mask map (shown in the right-most panel of Fig.\ 2). We assign a value of zero to all the pixels that fall within the region of a point source mask and assign a value of one otherwise. Each of these mask maps are constructed for individual sources and finally we co-add them to obtain the final mask map. The source map is divided by the mask map which adjusts for the multiple counting of pixels in the coadded source map and provides an average map of the sources. The left-most panel of Fig.\ 2 shows the raw stacked map of the L09 {\it Lowz} AGN sources in the soft band. Note that we do not apply any exposure time or effective area correction for this map and the point sources are unmasked. The left-middle panel shows the same map but now corrected for exposure time and effective area. However the point sources are still unmasked in these maps. We emphasize that the maps presented in the left-most and middle-left panels of Fig.\ 2 are for illustrative purposes only and have not been used in our actual analyses. In the middle-right panel we present the stacked map where the point sources are masked (excluding the central source). The right-most panel shows the stacked mask map. We divide the middle-right map by the right-most map to obtain the final maps.

Extracting surface brightness profiles requires subtraction of the background counts. We calculate the background counts using the average counts from a region that is larger than ten times the area of the PSF. We also require the region for background extraction to be sufficiently far away from our sources. This leads to the background counts being extracted from the annular region between 50 arcseconds and 70 arcseconds from the sources. The background count has been calculated using the stacked masked maps of our galaxies and AGN. We verified that changing the area of the region for background extraction does not affect the estimated background provided it is sufficiently far from the sources. The background count in the soft band is estimated to be $1.68 \times 10^{-10}$ cm$^{-2}$s$^{-1}$pixel$^{-1}$. For a $200$ ksec exposure and an effective area of $350$ cm$^{2}$ the background count is $\sim 0.012$ per pixel.  

To compare the surface brightness profiles of the AGN and galaxy maps we compute the spatial profiles of the mean photon counts of the two samples. We note that the total emission in the region is a combination of both diffuse emission in the ISM/ICM and the emission from the central nucleus. Since we are focused only on extended emission, we adopt a simple calibration to convert counts to flux. We assume that the average energy of each photon is equal to the average energy of the soft band ($1.25$ kev). This provides an order of magnitude estimate of the energy scales involved. The actual flux will depend on the spectrum and will be $\langle {\rm flux} \rangle /(1.25$ keV), which should be of order unity---which can be used to estimate the uncertainties in the flux calibration.

To separate the diffuse emission from the nuclear emission we construct a model PSF for our sources. The average model PSF is extracted from the calibrations in \citet{gaetzetal04}. We note that the wings of the PSF function are very extended and depend on the detector and the energy of the photon. A reasonable estimate of the PSF function comprising the core and a wing is 
\begin{eqnarray}
f(\theta)=\frac{A_{0}}{\left(1+\left(\frac{\theta}{\theta_{0}}\right)^{2}\right)^{\gamma/2}} + A_{1}\exp\left[-4\ln(2)\left(\frac{\theta}{\theta_{1}}\right)^{2}\right] \nonumber\\ + A_{2}\exp\left[-4\ln(2)\left(\frac{\theta}{\theta_{2}}\right)^{2}\right],
\end{eqnarray}
where the best-fit values of the parameters are given by \citet{gaetzetal04}. The energy of our photons lies in the range 0.5--2.0\,kev and hence we adopt the best-fit values of the PSF function corresponding to the energy scale 0.5--2.0\,kev. We emphasize that the PSF function for each individual galaxy and AGN source could be different from the mean function, but we use the mean PSF to be representative of our average stacked PSF. We further discuss this issue in \S 4. 

\begin{figure}[t]
\epsscale{1.00}
\plotone{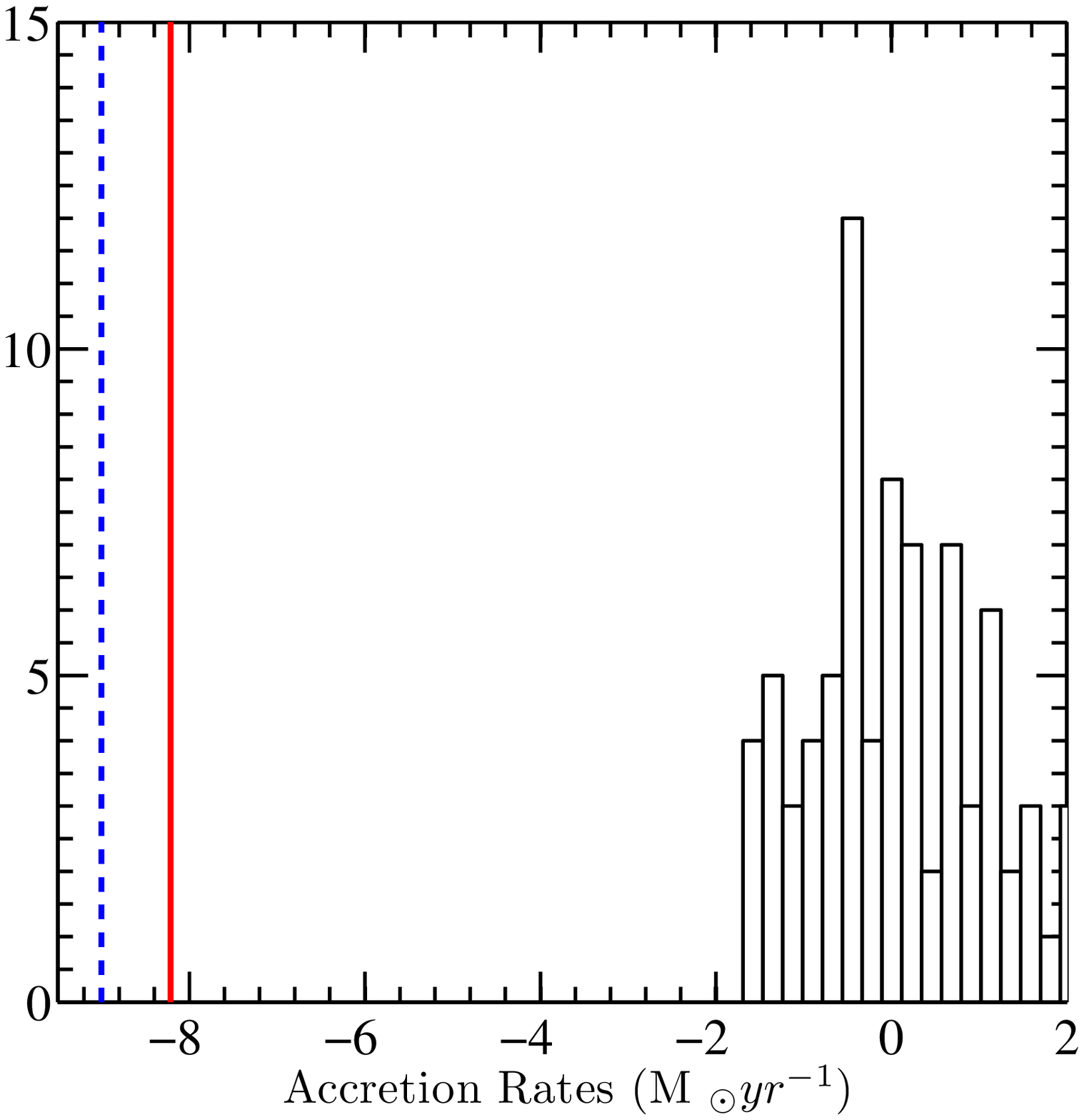}
\caption{Distribution of approximate accretion rates for our AGN sample. The red and the blue vertical lines represent the maximum energy difference between the AGN and the galaxy samples for the {\it Lowz} and the {\it Highz} samples, respectively. See \S 4 for the method of computing accretion rates. We note that the energy difference ($\Delta {\rm E}$) between our AGN and galaxy samples is $10^{-6}$--$10^{-7}$ times that of the typical accretion rates of our AGN. This is a few orders of magnitude lower than the assumed feedback fraction in theoretical studies, which is typically assumed to be $\sim 10^{-3}$--$10^{-4}$.} \label{fig-5}
\end{figure}

\section{Results}
We extract the surface brightness profiles from the average stacked maps. The X-ray profiles for the L09 and control samples are shown in Fig.\ 3. The units are in ${\rm cts\,s}^{-1}$arcsecond$^{-2}$. To obtain the profiles we estimate the mean count (per pixel) and the standard error (standard deviation/$\sqrt N$: N being the total number of pixels in each annulus) on the mean at each annulus. The left-hand and the right-hand panels represent the profiles for the {\it Lowz} and {\it Highz} samples, respectively. The blue dashed line and the green solid line in each panel shows the average PSF profiles (discussed in \S 3) for our {\it Lowz} and {\it Highz} AGN and galaxy sources, respectively. We normalize the PSF profiles to the peak emission from our galaxies and AGN. We note that in both cases we detect significant excesses ($\sim$ 1--5 $\sigma$) at scales in the range 1--10\,\arcsec, and that these excesses are significantly greater than the PSF wings. 

We emphasize that a key interpretation of our result lies in our ability to characterize the PSFs of our sources. As mentioned previously, the PSF profile described in Eq.\ 1 is an approximate expression of the PSF and is assumed to be an average PSF of our AGN and galaxy sources. In practice, though, the PSF of different sources would vary differently based on their position on the ACIS detector. L09 uses the MARX simulator (Wise et al.\ 2003) to characterize the Enclosed Energy Fraction (EEF) aperture size for our AGN sources. L09 also provides a look-up table from which the PSFs of the galaxies have been extracted. We note that the majority ($\sim$ 80--85 \%) of our AGN and galaxy sources have a 95\% EER at $<$ 5\arcsec. However there are a few sources that have  $5 \leq (95\%$ EER) $\leq 8$\arcsec. It is possible that we have PSF contamination from these few sources, and that this can compensate for a small fraction of the excess seen in Fig.\ 3. Note, though, that the same PSF profile cannot explain the {\em shape} of both the AGN and the galaxy profiles.  

From clustering measurements and from studies of environments it has been shown that X-ray bright AGN tend to reside in more massive halos \citep[e.g.,][]{gillietal05, monterodortaetal09,coiletal09, allevatoetal11, richardsonetal13}. The typical host halo mass scale of these AGN is $\sim 10^{13}M_{\odot}$ (group scale halos; \citealt{coiletal09}). This would imply a somewhat more extended emission around them due to the presence of the group potential-well, as is seen in the present case. However if galaxy groups are removed from the sample the host halo mass drops to $\sim 10^{12.5}M_{\odot}$ for our sample \citep{mountrichasetal13}. The effect will be similar for the galaxy control sample which has been matched in color and luminosity with the AGN host galaxies and thus should have similar environments to AGN host galaxies (see the discussion in \S 2.2). Thus, differences in the profiles between AGN and galaxies due to differences in environment will be minimal in this case.  

To quantify the differences in the counts from Fig.\ 3 we adopt the following technique. In Fig.\ 4 we plot the difference in excess photons between AGN and galaxies as a function of angular scale. 
\begin{eqnarray}
\alpha (\theta) &=& ({\rm count_{agn}}(\theta)- {\rm PSF_{agn}}(\theta)) - ({\rm count_{gal}}(\theta) -{\rm PSF_{gal}}(\theta)), \nonumber\\
\sigma_{{\rm diff}}(\theta) &=& \sqrt {\sigma_{{\rm agn}}^{2}(\theta) + \sigma_{{\rm gal}}^{2}}(\theta),  \nonumber\\ 
\end{eqnarray}
where ${\rm count_{agn}}$ and ${\rm count_{gal}}$ are the mean counts in each annulus (background subtracted), ${\rm PSF_{agn}}$ and ${\rm PSF_{gal}}$ are the PSF contributions, and $\sigma_{{\rm agn}}$ and $\sigma_{{\rm gal}}$ are the errors on mean counts in each annulus for the AGN and the galaxy samples, respectively. Positive (negative) numbers imply an excess (deficit) of X-ray flux in the AGN case. Black squares, and red circles represent the difference for the {\it Lowz} and {\it Highz} sample. In the left-hand panel of Fig.\ 4 we observe a huge excess from the AGN. However as discussed above there could be sources with larger PSFs and a very small fraction of the PSF can contaminate both the AGN and the galaxy signal. To better understand the nature of the excess, we plot the excess with reference to the PSF of the AGN and galaxies at that particular angular scale in the right panel of Fig.\ 4. We note that in the case of galaxies we observe more excess compared to its PSF than the AGN. This tentatively suggests that there is less hot gas in the vicinity of the AGN compared to normal galaxies. We further discuss this result in \S 5. 

We wish to stress that the interpretation of the value of $\alpha$ in Eq.\ 2 relies on our use of the average PSF of the stacked profile to normalize to the observed signal.  We have considered the average PSF of the stacked profile while normalizing it to the observed signal. The PSF profiles have been normalized to the peak value of the average stacked AGN emission and galaxy emission respectively (the leftmost black and red points in Fig.\ 3). The reason we normalize it to the peak (or rather the central-most) value is because we can easily assume that at the smallest scales we are unable to differentiate between extended emission and point-like emission and it would be safe to assume that 100\% of the emission is from the point source. As mentioned before we note that L09 values of the EER are comparable to our values with some small scatter. 

In the paper, we showed that the excess relative to the normalized PSF is higher in normal galaxies compared to AGN host galaxies. We argue that, a very plausible way to explain this observation is to interpret this excess as larger amount of extended emission in normal galaxies compared to AGN host galaxies at those particular angular scales.  Now, of course the PSF normalization for AGN is larger since they are the bright point sources detected in the AEGIS-X survey amd that is manifested in the overall higher counts for the AGN compared to galaxies (see the y axis of Fig.\ 3.  The black points are about 2 orders of magnitude larger than the AGN). 

The apparent positive and negative values are not related to the size of the PSF (which are similar for both AGN and normal galaxies), but are related to how much excess we get after subtracting the PSF. In other words alpha in the right panel of Fig.\ 4 denotes the difference in counts between AGN and galaxies at a given scale, when the PSF gets subtracted from both of them.

From Eq.\ 2 we can compute the approximate energy difference between AGN and galaxies. The maximum energy difference is given as $\Delta {\rm E}$ (ergs s$^{-1}$)= $2\pi {\theta {\rm d}\theta} \alpha(\theta){\rm d}_{{\rm L}}^{2}$, where d$_{{\rm L}}$ is the luminosity distance. To obtain luminosity distances we use the median redshift of the two samples. The median redshifts for the {\it Lowz} and {\it Highz} samples are $0.51$ and $0.86$ respectively. This gives a maximum energy difference of $3.5\times10^{38}$ ergs s$^{-1}$, and $5.7\times10^{37}$ ergs s$^{-1}$ for the {\it Lowz} and the {\it Highz} sample, respectively. Many theoretical models mostly assume feedback energy to be a fixed fraction of the accreted mass energy of the black hole, allowing more feedback from more luminous AGN \citep[e.g.,][]{dimatteoetal05}. We thus compare the approximate accretion rates of our AGN  samples with the energy difference ($\Delta {\rm E}$) derived from our surface brightness profiles. 

Several cosmological and isolated simulations of AGN feedback assume the feedback energy to be a fixed fraction of the bolometric luminosity (or accretion rate thereof) of the AGN \citep[e.g.,][] {c&o01, shankaretal04, dimatteoetal05, hopkinsetal06, johanssonetal08}. To compare our results with these models we extracted the accretion rates of our AGN sample. The accretion rates are compiled as follows. The bolometric luminosities of our AGN sample were obtained from the X-ray luminosities using the bolometric corrections of \citet{marconietal04}. We then assumed $\eta=0.1$, a canonical efficiency for thin disk accretion \citep{s&s73} and computed the approximate accretion rates of the black holes corresponding to our AGN sample. The distribution of accretion rates is shown in Fig.\ 5. The red and the blue vertical lines refer, respectively, to the equivalent mass difference for the {\it Lowz} and {\it Highz} samples corresponding to the energy differences ($\Delta {\rm E}$) derived from Fig.\ 3 and Eq.\ 2. We note that the energy difference ($\Delta {\rm E}$) between our AGN and galaxy samples is $10^{-6}$--$10^{-7}$ of that of the typical accretion rates of our AGN. This is a few orders of magnitude lower than the assumed feedback fraction in theoretical studies (mentioned above), which is typically assumed to be $\sim 10^{-3}$--$10^{-4}$.

\section{Discussion of Results}

As discussed before, models of AGN feedback suggest that there should be a relationship between the outflows/energy from an AGN and the density and temperature distribution of gas in galaxies and clusters. However the magnitude and the scale of influence remain mostly uncertain. For example, studies suggest that the scale of influence of AGN feedback (where observable signatures are prominent) can vary from a few tens of kpc \citep[e.g.,][]{pelligrinietal12} to a few hundred kpc \citep[e.g.,][]{chatterjeeetal08} to a few Mpc \citep[e.g.,][]{scannapiecoetal08} depending on the nature of the observable. 

Based on simulations of an isolated elliptical galaxy (with a $B$-band luminosity of $L_{B} = 5 \times 10^{10}{\rm L}_{B {\odot}}$) \citet{pelligrinietal12} found that the X-ray surface brightness profiles in galaxies are significantly different if the effects of feedback from an AGN are included. Differences in the surface brightness profile are evident even at radii beyond 50 kpc. We note that the scale of influence as predicted by (\citealt{pelligrinietal12}; a few kpc, sub-arcsecond for our purposes) where features in the surface brightness profiles due to the nuclear outburst can be visible, is well within the PSF scales of our AGN sources. Without decomposing the observed emission into PSF-like and extended components, we cannot distinguish models based on features in the surface brightness profile at small radius. However large scale differences ($\sim 50$ kpc) due to AGN feedback are likely to show up in the X-ray profiles. We search for such differences in the surface brightness profile. 

\citet{gasparietal11} use 3D adaptive mesh refinement simulations to carefully constrain the effects of feedback in the hot ISM/ICM. The results show a central depression in the surface brightness profile. This is in accordance with a physical picture in which feedback from AGN is likely to disrupt the ISM/ICM gas and transport it to a larger length scale from the center of the galaxy/cluster. This appears as bumps and cavities at larger (far from the black hole) and smaller (closer to the black hole) length scales in the X-ray surface brightness profiles. \citet{gasparietal11} shows that the evacuation of gas can occur at scales of tens of kpc (in galaxies/groups) to hundreds of kpc (in clusters). Using a smoothed particle hydrodynamics simulation \citet{choietal13} shows that AGN feedback drives gas from the center of the galaxy and lowers the X-ray luminosity of the hot halo. 

As mentioned in \S 4 the characterization of the difference between AGN and galaxies is highly sensitive to our understanding of the true stacked PSF. Thus we are able to make only qualitative comparisons with theoretical studies. In Fig.\ 4 we show the difference in the excess flux between AGN and galaxies. In the left-hand panel we plot the statistic $\alpha$ (Eq.\ 2) as a function of angular scale. It shows that the AGN have excess flux over the galaxies at scales below 5\arcsec. We note that majority of our sources have a 95\% EER $\leq$ 5\arcsec. Thus it is difficult to quantify the true nature of this excess. We thus normalize the excess emission to the corresponding PSFs in Fig.\ 3 and show the normalized difference in the right panel of Fig.\ 4. This suggests that the extended emission in galaxies compared to the PSF is much higher than the case for AGN. This is qualitatively similar to theoretical studies in which it has been shown that AGN feedback drives the gas from the ISM/ICM out to scales of tens to hundreds of kpc, eventually lowering the X-ray luminosity of the hot ISM/ICM. 

We also note that different point source poplulations such as low mass X-ray binaries (LMXB) and stellar X-ray sources (ABs and CVs) will have an impact in characterizing the extended emission. Our galaxy and and AGN sample are mostly z $\sim 1$ objects. The LMXB population in these high redshift galaxies is not well characterized. We thus assume that if the stellar masses and luminosities of galaxies are correlated with the LMXB population we can assume that the LMXB population is similar in AGN and non AGN galaxies. However, we stress that this conclusion is subject to the assumption that the populations {\it are} identical.

In addition to the spatial scale of feedback, the magnitude of feedback is an important quantity that remains uncertain in the literature. We note that the maximum energy difference (the actual value is likely to be lower) that we detect between AGN and galaxies is $10^{-6}$--$10^{-7}$ times the typical bolometric luminosities of our AGN sources. This is few orders of magnitude lower than the assumed feedback fraction in the literature. We clarify that the lack of a major difference between AGN and control samples, does not exclude stronger feedback. It is likely that the AGN that are visible in the X-ray are essentially just a random subset of all galaxies: those that happen to be accreting strongly enough to be detected {\em now}; but the duty cycle is short. Thus all of the objects in the control sample may have had AGN feedback recently, too, but just happen to be currently ``turned off'', and hence have a somewhat similar net profile. We discuss the possibility of extending this work using alternative AGN samples in \S 5.1.
  
\subsection{Future Work}

Although we find substantial differences between the X-ray emission around galaxies and AGN, PSF contamination in our surface-brightness profile is a major source of uncertainty in characterizing the significance of this signal. Hence, a sample of black holes that have comparable accretion rates to that of our X-ray AGN, but are under-luminous in X-rays, will be an ideal sample for studying AGN-ISM/ICM interaction at these scales. Confusion due to the nuclear emission will be limited in this case, making it easier to extract any extended emission. Extended emission would still be expected if feedback is strongly correlated with accretion rate, as is assumed in many theoretical models. We propose to undertake these studies in a future paper. 

The theoretical paradigm of ``accretion rate dependent feedback'' could be examined by comparing the X-ray surface brightness profiles of highly accreting black holes---e.g., quasars, for which the feedback energy is expected to be higher if we assume the feedback energy to be directly proportional to the bolometric luminosity/accretion rates of the AGN---to that of normal galaxies with identical optical properties to quasar hosts. Since inference about the optical properties of the host galaxies of quasars is difficult, the best sample with which to conduct this measurement might be a population of optically obscured but IR-bright quasars. 

In addition, quasars may grow most in the obscured phase and hence would have the highest amount of feedback during this phase \citep{hopkinsetal08}. If this paradigm is correct we would expect to see the maximum effect of feedback on X-ray surface brightness profiles around obscured quasars. The success of this study will depend on the time-scale of feedback, compared to the typical lifetime of a luminous quasar ($\sim 10$Myr). Some theoretical models suggest that about 20--30\% of the quasar population undergo a rapid blowout phase \citep[e.g.,][]{hopkinsetal08}. In short, a promising future avenue might be to apply our analysis to a study of the effect of obscured quasars on the diffuse ISM.   

\section{Summary} 
In this work, we perform a stacking analysis of X-ray selected AGN in the AEGIS field in the redshift range 0.3--1.3 and compare their average surface brightness profile to the average surface brightness profile of galaxies in the same field. Our AGN and galaxies are matched in optical properties and redshift distributions. We tentatively detect extended emission in the ISM/ICM at a scale of 1--$10\arcsec$ for both accreting and non-accreting galaxies. The exact significance of the detection is sensitive to the true characterization of the PSF. To quantify the differences in the X-ray profiles between AGN and galaxies, we extract the mean-difference profile ($\alpha$ in Eq.\ 1). 

When normalized by the PSF, we note that galaxies tend to have more extended X-ray emission than AGN. This result is qualitatively similar to the predictions from theoretical simulations in which AGN feedback has been linked with evacuating gas from the center of the galaxy to larger scales. Since contamination of the {\it Chandra} PSF will generally be higher for the brightest X-ray sources, we propose that a sample of black holes that have comparable accretion rates to that of our X-ray AGN, but that are under-luminous in X-rays, will be an ideal sample for studying the effect of feedback on diffuse ISM/ICM gas.  We also suggest that obscured quasars might have more effect on X-ray surface brightness profiles if the amount of feedback energy is directly proportional to the accretion rate of the black hole, as assumed in several theoretical models.

\section*{Acknowledgments}
SC would like to thank Nico Cappelluti for some discussions on background counts in our maps. SC would also like to thank Silvia Pellegrini, Thorsten Naab, and Massimo Gaspari for some useful discussions about their AGN feedback simulations and Terry Gaetz about issues with the Chandra PSF. We would also like to thank the referee for several useful suggestions which helped in significant improvement of the draft. SC thanks Daisuke Nagai for several useful discussions and acknowledges computational resources and support from YCAA. 

SC and JAN were partially supported by NSF grant AST-0806732 at the University of Pittsburgh. SC and ADM were partially supported by the National Science Foundation through grant number 1211112, and by the National Aeronautics Space Administration (NASA) through ADAP award NNX12AE38G and Chandra Award Number AR0-11018C issued by the Chandra X-ray Observatory Center, which is operated by the Smithsonian Astrophysical Observatory for and on behalf of NASA under contract NAS8-03060. ALC acknowledges support from NSF CAREER award AST-1055081.


      
\bibliography{mybib}{}

\begin{thebibliography}{67}
\expandafter\ifx\csname natexlab\endcsname\relax\def\natexlab#1{#1}\fi

\bibitem[{{Allevato} {et~al.}(2011){Allevato}, {Finoguenov}, {Cappelluti},
  {Miyaji}, {Hasinger}, {Salvato}, {Brusa}, {Gilli}, {Zamorani}, {Shankar},
  {James}, {McCracken}, {Bongiorno}, {Merloni}, {Peacock}, {Silverman}, \&
  {Comastri}}]{allevatoetal11}
{Allevato}, V., {et~al.} 2011, \apj, 736, 99

\bibitem[{{Arnaud} \& {Evrard}(1999)}]{a&e99}
{Arnaud}, M., \& {Evrard}, A.~E. 1999, \mnras, 305, 631

\bibitem[{{Battaglia} {et~al.}(2010){Battaglia}, {Bond}, {Pfrommer}, {Sievers},
  \& {Sijacki}}]{battagliaetal10}
{Battaglia}, N., {Bond}, J.~R., {Pfrommer}, C., {Sievers}, J.~L., \& {Sijacki},
  D. 2010, \apj, 725, 91

\bibitem[{{Bell} \& {de Jong}(2001)}]{b&dejong01}
{Bell}, E.~F., \& {de Jong}, R.~S. 2001, \apj, 550, 212

\bibitem[{{Bhattacharya} {et~al.}(2008){Bhattacharya}, {Di Matteo}, \&
  {Kosowsky}}]{bhattacharyaetal08}
{Bhattacharya}, S., {Di Matteo}, T., \& {Kosowsky}, A. 2008, \mnras, 389, 34

\bibitem[{{B{\^i}rzan} {et~al.}(2004){B{\^i}rzan}, {Rafferty}, {McNamara},
  {Wise}, \& {Nulsen}}]{birzanetal04}
{B{\^i}rzan}, L., {Rafferty}, D.~A., {McNamara}, B.~R., {Wise}, M.~W., \&
  {Nulsen}, P.~E.~J. 2004, \apj, 607, 800

\bibitem[{{Booth} \& {Schaye}(2009)}]{b&s09}
{Booth}, C.~M., \& {Schaye}, J. 2009, \mnras, 398, 53

\bibitem[{{Cattaneo} {et~al.}(2006){Cattaneo}, {Dekel}, {Devriendt},
  {Guiderdoni}, \& {Blaizot}}]{cattaneoetal06}
{Cattaneo}, A., {Dekel}, A., {Devriendt}, J., {Guiderdoni}, B., \& {Blaizot},
  J. 2006, \mnras, 370, 1651

\bibitem[{{Chatterjee} {et~al.}(2008){Chatterjee}, {Di Matteo}, {Kosowsky}, \&
  {Pelupessy}}]{chatterjeeetal08}
{Chatterjee}, S., {Di Matteo}, T., {Kosowsky}, A., \& {Pelupessy}, I. 2008,
  \mnras, 390, 535

\bibitem[{{Chatterjee} {et~al.}(2010){Chatterjee}, {Ho}, {Newman}, \&
  {Kosowsky}}]{chatterjeeetal10}
{Chatterjee}, S., {Ho}, S., {Newman}, J.~A., \& {Kosowsky}, A. 2010, \apj, 720,
  299

\bibitem[{{Chatterjee} \& {Kosowsky}(2007)}]{c&k07}
{Chatterjee}, S., \& {Kosowsky}, A. 2007, \apjl, 661, L113

\bibitem[{{Choi} {et~al.}(2013){Choi}, {Naab}, {Ostriker}, {Johansson}, \&
  {Moster}}]{choietal13}
{Choi}, E., {Naab}, T., {Ostriker}, J.~P., {Johansson}, P.~H., \& {Moster},
  B.~P. 2013, ArXiv e-prints

\bibitem[{{Choi} {et~al.}(2012){Choi}, {Ostriker}, {Naab}, \&
  {Johansson}}]{choietal12}
{Choi}, E., {Ostriker}, J.~P., {Naab}, T., \& {Johansson}, P.~H. 2012, \apj,
  754, 125

\bibitem[{{Ciotti} \& {Ostriker}(2001)}]{c&o01}
{Ciotti}, L., \& {Ostriker}, J.~P. 2001, \apj, 551, 131

\bibitem[{{Ciotti} \& {Ostriker}(2007)}]{c&o07}
---. 2007, \apj, 665, 1038

\bibitem[{{Coil} {et~al.}(2004){Coil}, {Newman}, {Kaiser}, {Davis}, {Ma},
  {Kocevski}, \& {Koo}}]{coiletal04}
{Coil}, A.~L., {Newman}, J.~A., {Kaiser}, N., {Davis}, M., {Ma}, C.-P.,
  {Kocevski}, D.~D., \& {Koo}, D.~C. 2004, \apj, 617, 765

\bibitem[{{Coil} {et~al.}(2009){Coil}, {Georgakakis}, {Newman}, {Cooper},
  {Croton}, {Davis}, {Koo}, {Laird}, {Nandra}, {Weiner}, {Willmer}, \&
  {Yan}}]{coiletal09}
{Coil}, A.~L., {et~al.} 2009, \apj, 701, 1484

\bibitem[{{Cooper} {et~al.}(2009){Cooper}, {Newman}, \& {Yan}}]{cooperetal09}
{Cooper}, M.~C., {Newman}, J.~A., \& {Yan}, R. 2009, \apj, 704, 687

\bibitem[{{Croton} {et~al.}(2006){Croton}, {Springel}, {White}, {De Lucia},
  {Frenk}, {Gao}, {Jenkins}, {Kauffmann}, {Navarro}, \&
  {Yoshida}}]{crotonetal06}
{Croton}, D.~J., {et~al.} 2006, \mnras, 365, 11

\bibitem[{{Davis} {et~al.}(2003){Davis}, {Faber}, {Newman}, {Phillips},
  {Ellis}, {Steidel}, {Conselice}, {Coil}, {Finkbeiner}, {Koo}, {Guhathakurta},
  {Weiner}, {Schiavon}, {Willmer}, {Kaiser}, {Luppino}, {Wirth}, {Connolly},
  {Eisenhardt}, {Cooper}, \& {Gerke}}]{davisetal03}
{Davis}, M., {et~al.} 2003, 4834, 161

\bibitem[{{Davis} {et~al.}(2007){Davis}, {Guhathakurta}, {Konidaris}, {Newman},
  {Ashby}, {Biggs}, {Barmby}, {Bundy}, {Chapman}, {Coil}, {Conselice},
  {Cooper}, {Croton}, {Eisenhardt}, {Ellis}, {Faber}, {Fang}, {Fazio},
  {Georgakakis}, {Gerke}, {Goss}, {Gwyn}, {Harker}, {Hopkins}, {Huang},
  {Ivison}, {Kassin}, {Kirby}, {Koekemoer}, {Koo}, {Laird}, {Le Floc'h}, {Lin},
  {Lotz}, {Marshall}, {Martin}, {Metevier}, {Moustakas}, {Nandra}, {Noeske},
  {Papovich}, {Phillips}, {Rich}, {Rieke}, {Rigopoulou}, {Salim},
  {Schiminovich}, {Simard}, {Smail}, {Small}, {Weiner}, {Willmer}, {Willner},
  {Wilson}, {Wright}, \& {Yan}}]{davisetal07}
---. 2007, \apjl, 660, L1

\bibitem[{{Di Matteo} {et~al.}(2005){Di Matteo}, {Springel}, \&
  {Hernquist}}]{dimatteoetal05}
{Di Matteo}, T., {Springel}, V., \& {Hernquist}, L. 2005, \nat, 433, 604

\bibitem[{{Dong} {et~al.}(2010){Dong}, {Rasmussen}, \& {Mulchaey}}]{dongetal10}
{Dong}, R., {Rasmussen}, J., \& {Mulchaey}, J.~S. 2010, \apj, 712, 883

\bibitem[{{Dunn} \& {Fabian}(2006)}]{d&f06}
{Dunn}, R.~J.~H., \& {Fabian}, A.~C. 2006, \mnras, 373, 959

\bibitem[{{Gaetz} {et~al.}(2004){Gaetz}, {Edgar}, {Jerius}, {Zhao}, \&
  {Smith}}]{gaetzetal04}
{Gaetz}, T.~J., {Edgar}, R.~J., {Jerius}, D.~H., {Zhao}, P., \& {Smith}, R.~K.
  2004, in Society of Photo-Optical Instrumentation Engineers (SPIE) Conference
  Series, Vol. 5165, Society of Photo-Optical Instrumentation Engineers (SPIE)
  Conference Series, ed. K.~A. {Flanagan} \& O.~H.~W. {Siegmund}, 411--422

\bibitem[{{Gaspari} {et~al.}(2011){Gaspari}, {Brighenti}, {D'Ercole}, \&
  {Melioli}}]{gasparietal11}
{Gaspari}, M., {Brighenti}, F., {D'Ercole}, A., \& {Melioli}, C. 2011, \mnras,
  415, 1549

\bibitem[{{Gaspari} {et~al.}(2012){Gaspari}, {Brighenti}, \&
  {Temi}}]{gasparietal12}
{Gaspari}, M., {Brighenti}, F., \& {Temi}, P. 2012, \mnras, 424, 190

\bibitem[{{Gebhardt} {et~al.}(2000){Gebhardt}, {Bender}, {Bower}, {Dressler},
  {Faber}, {Filippenko}, {Green}, {Grillmair}, {Ho}, {Kormendy}, {Lauer},
  {Magorrian}, {Pinkney}, {Richstone}, \& {Tremaine}}]{gebhardtetal00}
{Gebhardt}, K., {et~al.} 2000, \apjl, 539, L13

\bibitem[{{Georgakakis} {et~al.}(2007){Georgakakis}, {Nandra}, {Laird},
  {Cooper}, {Gerke}, {Newman}, {Croton}, {Davis}, {Faber}, \&
  {Coil}}]{georgakakisetal07}
{Georgakakis}, A., {et~al.} 2007, \apjl, 660, L15

\bibitem[{{Gilli} {et~al.}(2005){Gilli}, {Daddi}, {Zamorani}, {Tozzi},
  {Borgani}, {Bergeron}, {Giacconi}, {Hasinger}, {Mainieri}, {Norman},
  {Rosati}, {Szokoly}, \& {Zheng}}]{gillietal05}
{Gilli}, R., {et~al.} 2005, \aap, 430, 811

\bibitem[{{Giodini} {et~al.}(2010){Giodini}, {Smol{\v c}i{\'c}}, {Finoguenov},
  {Boehringer}, {B{\^i}rzan}, {Zamorani}, {Oklop{\v c}i{\'c}}, {Pierini},
  {Pratt}, {Schinnerer}, {Massey}, {Koekemoer}, {Salvato}, {Sanders},
  {Kartaltepe}, \& {Thompson}}]{giodinietal10}
{Giodini}, S., {et~al.} 2010, \apj, 714, 218

\bibitem[{{Gitti} {et~al.}(2012){Gitti}, {Brighenti}, \&
  {McNamara}}]{gittietal12}
{Gitti}, M., {Brighenti}, F., \& {McNamara}, B.~R. 2012, Advances in Astronomy,
  2012

\bibitem[{{Hopkins} {et~al.}(2008){Hopkins}, {Hernquist}, {Cox}, \& {Kere{\v
  s}}}]{hopkinsetal08}
{Hopkins}, P.~F., {Hernquist}, L., {Cox}, T.~J., \& {Kere{\v s}}, D. 2008,
  \apjs, 175, 356

\bibitem[{{Hopkins} {et~al.}(2006){Hopkins}, {Robertson}, {Krause},
  {Hernquist}, \& {Cox}}]{hopkinsetal06}
{Hopkins}, P.~F., {Robertson}, B., {Krause}, E., {Hernquist}, L., \& {Cox},
  T.~J. 2006, \apj, 652, 107

\bibitem[{{Johansson} {et~al.}(2008){Johansson}, {Naab}, \&
  {Burkert}}]{johanssonetal08}
{Johansson}, P.~H., {Naab}, T., \& {Burkert}, A. 2008, Astronomische
  Nachrichten, 329, 956

\bibitem[{{Johnson} {et~al.}(2009){Johnson}, {Ponman}, \&
  {Finoguenov}}]{johnsonetal09}
{Johnson}, R., {Ponman}, T.~J., \& {Finoguenov}, A. 2009, \mnras, 395, 1287

\bibitem[{{Kauffmann} \& {Haehnelt}(2000)}]{k&h00}
{Kauffmann}, G., \& {Haehnelt}, M. 2000, \mnras, 311, 576

\bibitem[{{Laird} {et~al.}(2009){Laird}, {Nandra}, {Georgakakis}, {Aird},
  {Barmby}, {Conselice}, {Coil}, {Davis}, {Faber}, {Fazio}, {Guhathakurta},
  {Koo}, {Sarajedini}, \& {Willmer}}]{lairdetal09}
{Laird}, E.~S., {et~al.} 2009, \apjs, 180, 102

\bibitem[{{Lapi} {et~al.}(2006){Lapi}, {Shankar}, {Mao}, {Granato}, {Silva},
  {De Zotti}, \& {Danese}}]{lapietal06}
{Lapi}, A., {Shankar}, F., {Mao}, J., {Granato}, G.~L., {Silva}, L., {De
  Zotti}, G., \& {Danese}, L. 2006, \apj, 650, 42

\bibitem[{{Marconi} {et~al.}(2004){Marconi}, {Risaliti}, {Gilli}, {Hunt},
  {Maiolino}, \& {Salvati}}]{marconietal04}
{Marconi}, A., {Risaliti}, G., {Gilli}, R., {Hunt}, L.~K., {Maiolino}, R., \&
  {Salvati}, M. 2004, \mnras, 351, 169

\bibitem[{{McNamara} \& {Nulsen}(2007)}]{m&n07}
{McNamara}, B.~R., \& {Nulsen}, P.~E.~J. 2007, \araa, 45, 117

\bibitem[{{Merritt} \& {Ferrarese}(2001)}]{m&f01}
{Merritt}, D., \& {Ferrarese}, L. 2001, \apj, 547, 140

\bibitem[{{Montero-Dorta} {et~al.}(2009){Montero-Dorta}, {Croton}, {Yan},
  {Cooper}, {Newman}, {Georgakakis}, {Prada}, {Davis}, {Nandra}, \&
  {Coil}}]{monterodortaetal09}
{Montero-Dorta}, A.~D., {et~al.} 2009, \mnras, 392, 125

\bibitem[{{Mountrichas} {et~al.}(2013){Mountrichas}, {Georgakakis},
  {Finoguenov}, {Erfanianfar}, {Cooper}, {Coil}, {Laird}, {Nandra}, \&
  {Newman}}]{mountrichasetal13}
{Mountrichas}, G., {et~al.} 2013, \mnras, 430, 661

\bibitem[{{Nandra} {et~al.}(2005){Nandra}, {Laird}, {Adelberger}, {Gardner},
  {Mushotzky}, {Rhodes}, {Steidel}, {Teplitz}, \& {Arnaud}}]{nandraetal05}
{Nandra}, K., {et~al.} 2005, \mnras, 356, 568

\bibitem[{{Nandra} {et~al.}(2007){Nandra}, {Georgakakis}, {Willmer}, {Cooper},
  {Croton}, {Davis}, {Faber}, {Koo}, {Laird}, \& {Newman}}]{nandraetal07}
---. 2007, \apjl, 660, L11

\bibitem[{{Nath} \& {Roychowdhury}(2002)}]{n&r02}
{Nath}, B.~B., \& {Roychowdhury}, S. 2002, \mnras, 333, 145

\bibitem[{{Newman} {et~al.}(2013){Newman}, {Cooper}, {Davis}, {Faber}, {Coil},
  {Guhathakurta}, {Koo}, {Phillips}, {Conroy}, {Dutton}, {Finkbeiner}, {Gerke},
  {Rosario}, {Weiner}, {Willmer}, {Yan}, {Harker}, {Kassin}, {Konidaris},
  {Lai}, {Madgwick}, {Noeske}, {Wirth}, {Connolly}, {Kaiser}, {Kirby},
  {Lemaux}, {Lin}, {Lotz}, {Luppino}, {Marinoni}, {Matthews}, {Metevier}, \&
  {Schiavon}}]{newmanetal13}
{Newman}, J.~A., {et~al.} 2013, \apjs, 208, 5

\bibitem[{{Novak} {et~al.}(2011){Novak}, {Ostriker}, \& {Ciotti}}]{novaketal11}
{Novak}, G.~S., {Ostriker}, J.~P., \& {Ciotti}, L. 2011, \apj, 737, 26

\bibitem[{{Nulsen} {et~al.}(2005){Nulsen}, {Hambrick}, {McNamara}, {Rafferty},
  {Birzan}, {Wise}, \& {David}}]{nulsenetal05}
{Nulsen}, P.~E.~J., {Hambrick}, D.~C., {McNamara}, B.~R., {Rafferty}, D.,
  {Birzan}, L., {Wise}, M.~W., \& {David}, L.~P. 2005, \apjl, 625, L9

\bibitem[{{O'Sullivan} {et~al.}(2010){O'Sullivan}, {Giacintucci}, {David},
  {Vrtilek}, \& {Raychaudhury}}]{osullivanetal10}
{O'Sullivan}, E., {Giacintucci}, S., {David}, L.~P., {Vrtilek}, J.~M., \&
  {Raychaudhury}, S. 2010, \mnras, 407, 321

\bibitem[{{Pellegrini} {et~al.}(2012){Pellegrini}, {Ciotti}, \&
  {Ostriker}}]{pelligrinietal12}
{Pellegrini}, S., {Ciotti}, L., \& {Ostriker}, J.~P. 2012, \apj, 744, 21

\bibitem[{{Peterson} \& {Fabian}(2006)}]{p&f06}
{Peterson}, J.~R., \& {Fabian}, A.~C. 2006, \physrep, 427, 1

\bibitem[{{Puchwein} {et~al.}(2010){Puchwein}, {Springel}, {Sijacki}, \&
  {Dolag}}]{puchweinetal10}
{Puchwein}, E., {Springel}, V., {Sijacki}, D., \& {Dolag}, K. 2010, \mnras,
  406, 936

\bibitem[{{Randall} {et~al.}(2011){Randall}, {Forman}, {Giacintucci}, {Nulsen},
  {Sun}, {Jones}, {Churazov}, {David}, {Kraft}, {Donahue}, {Blanton},
  {Simionescu}, \& {Werner}}]{randalletal11}
{Randall}, S.~W., {et~al.} 2011, \apj, 726, 86

\bibitem[{{Richardson} {et~al.}(2013){Richardson}, {Chatterjee}, {Zheng},
  {Myers}, \& {Hickox}}]{richardsonetal13}
{Richardson}, J., {Chatterjee}, S., {Zheng}, Z., {Myers}, A.~D., \& {Hickox},
  R. 2013, \apj, 774, 143

\bibitem[{{Scannapieco} \& {Oh}(2004)}]{s&o04}
{Scannapieco}, E., \& {Oh}, S.~P. 2004, \apj, 608, 62

\bibitem[{{Scannapieco} {et~al.}(2008){Scannapieco}, {Thacker}, \&
  {Couchman}}]{scannapiecoetal08}
{Scannapieco}, E., {Thacker}, R.~J., \& {Couchman}, H.~M.~P. 2008, \apj, 678,
  674

\bibitem[{{Schawinski} {et~al.}(2007){Schawinski}, {Thomas}, {Sarzi},
  {Maraston}, {Kaviraj}, {Joo}, {Yi}, \& {Silk}}]{schawinskietal07}
{Schawinski}, K., {Thomas}, D., {Sarzi}, M., {Maraston}, C., {Kaviraj}, S.,
  {Joo}, S.-J., {Yi}, S.~K., \& {Silk}, J. 2007, \mnras, 382, 1415

\bibitem[{{Shakura} \& {Sunyaev}(1973)}]{s&s73}
{Shakura}, N.~I., \& {Sunyaev}, R.~A. 1973, \aap, 24, 337

\bibitem[{{Shankar} {et~al.}(2004){Shankar}, {Salucci}, {Granato}, {De Zotti},
  \& {Danese}}]{shankaretal04}
{Shankar}, F., {Salucci}, P., {Granato}, G.~L., {De Zotti}, G., \& {Danese}, L.
  2004, \mnras, 354, 1020

\bibitem[{{Sijacki} {et~al.}(2007){Sijacki}, {Springel}, {Di Matteo}, \&
  {Hernquist}}]{sijackietal07}
{Sijacki}, D., {Springel}, V., {Di Matteo}, T., \& {Hernquist}, L. 2007,
  \mnras, 380, 877

\bibitem[{{Sunyaev} \& {Zeldovich}(1972)}]{s&z72}
{Sunyaev}, R.~A., \& {Zeldovich}, Y.~B. 1972, Comments on Astrophysics and
  Space Physics, 4, 173

\bibitem[{{Teyssier} {et~al.}(2011){Teyssier}, {Moore}, {Martizzi}, {Dubois},
  \& {Mayer}}]{teyssieretal11}
{Teyssier}, R., {Moore}, B., {Martizzi}, D., {Dubois}, Y., \& {Mayer}, L. 2011,
  \mnras, 618

\bibitem[{{Thacker} {et~al.}(2009){Thacker}, {Scannapieco}, {Couchman}, \&
  {Richardson}}]{thackeretal09}
{Thacker}, R.~J., {Scannapieco}, E., {Couchman}, H.~M.~P., \& {Richardson}, M.
  2009, \apj, 693, 552

\bibitem[{{Tremaine} {et~al.}(2002){Tremaine}, {Gebhardt}, {Bender}, {Bower},
  {Dressler}, {Faber}, {Filippenko}, {Green}, {Grillmair}, {Ho}, {Kormendy},
  {Lauer}, {Magorrian}, {Pinkney}, \& {Richstone}}]{tremaineetal02}
{Tremaine}, S., {et~al.} 2002, \apj, 574, 740

\bibitem[{{Wyithe} \& {Loeb}(2003)}]{w&l03}
{Wyithe}, J.~S.~B., \& {Loeb}, A. 2003, \apj, 595, 614

\end{thebibliography}
\bibliographystyle{apj}
\end{document}